# A Review of Salam Phase Transition in Protein Amino Acids: Implication for Biomolecular Homochirality

Bai Fan[1]   Wang Wenqing[2*]   Min Wei[2]

1. Department of Physics, Peking University, Beijing, 100871, China

2. Department of Chemistry, Peking University, Beijing, 100871, China

**Dedicated to Professor Abdus Salam**

*ABSTRACT*   The origin of chirality, closely related to the evolution of life on the earth, has long been debated. In 1991, Abdus Salam suggested a novel approach to achieve biomolecular homochirality by a phase transition. In his subsequent publication, he predicted that this phase transition could eventually change D-amino acids to L-amino acids as $C_\alpha$-H bond would break and H atom became a superconductive atom. Since many experiments denied the configuration change in amino acids, Salam hypothesis aroused suspicion. This paper is aimed to provide direct experimental evidence of a phase transition in alanine, valine single crystals but deny the configuration change of D- to L-enantiomers. New views on Salam phase transition are presented to recover its great importance in the origin of homochirality.

## ORIGIN OF BIOMOLECULAR HOMOCHIRALITY

In 1848, Pasteur first observed that crystals formed from a salt of tartaric acid were of two types, each one a mirror image of the other. He also discovered that solutions of each type of these crystals rotated polarized light in opposite directions. Now we all know the two types of crystals correspond to mirror-symmetric arrangements of atoms in the molecule. The origin of the chiral purity (L-amino acids and D-sugar are dominant in living organisms) has been a constant preoccupation in biology, but the problem is not solved yet. Among those hypotheses concerning this problem, two main directions are obvious to be summarized as whether homochirality is the consequence of a chance process or is induced by an asymmetric physical force. Competing autocatalysis and the symmetry fractures during crystallization describe the occurrence of homochirality in biomolecules as accidental. The basic concept of the alternative "determined mechanism" is an external physical interaction, which is able to produce an enantiomerically enhanced product from an optically inactive starting material, or which utilizes chirality already intrinsic to the molecule.

---

[*] Author for all correspondence: wangwq@sun.ihep.ac.cn

In 1957, Lee and Yang discovered the parity violation effect in weak interactions [1]. Then it became probably the most promising physical effect for inducing effects on a macroscopic scale. Soon after the unification of weak and electromagnetic interactions, it was realized that different enantiomers would possess different physical properties, due to the existence of neutral weak currents. Ab initio calculation were carried out for various amino-acids in aqueous solutions showing parity violating energy differences of the order of $10^{-17}$ kT; this corresponds to an excess of approximately $10^{-17}$ for the molecules of lower energy, a very small number indeed [2]. Some orders of magnitude can be gained if one considers instead some other (heavier) biomolecules, but an immense gap is still to be bridged before a connection is established with biomolecular homochirality.

Before Salam phase transition theory, other hypotheses had been suggested to explain homochirality. One of the principal difficulties in understanding the asymmetry of biomolecules involves the amplification mechanism that would boost a primordial small physical effect. All suggestions concerning amplification mechanism should consequently be considered seriously:

1. Amplification of the Vester-Ulbricht processes, namely, amplification mechanisms that may apply to the interaction of polarized particles with biomolecules [3]. This hypothesis postulated that when the longitudinally polarized electrons produced by parity violation during $\beta$-decay impinged on matter to generate circularly polarized Bremsstrahlung photons, these in turn could be absorbed by organic substrates, thereby inducing stereoselective synthetic or degradative reactions of sorts known to be brought about by circularly polarized light. The overall result would be the abiotic generation of chiral organic products.

2. Amplification of the Yamagata processes, namely, processes in which the breakdown of the discrete P symmetry is considered as a possible source of biochirality [4]. The accumulation principle suggested by Yamagata conjectures that a small difference $\varepsilon$ in the rate for L-molecules to form pure L-polymer vs. the rate for D-molecules to form pure D-polymer will eventually lead to $\varepsilon$ n-times more L polymer of n members than D-polymer of n members from racemic monomer solution.

3. Amplification such as the Kondepudi-Nelson catastrophic mechanism [5]. According to this, the small electroweak energy difference between the enantiomers is sufficient to determine which of the two single-enantiomer reaction sequences was adopted at the metastable stage in the flow reactor that is off the thermodynamic equilibrium

It had been suggested that Bose-Einstein condensation is a novel approach to achieve homochirality. This theory is suggested by Abdus Salam, which we will discuss in detail in the following section of this paper.

## SALAM PHASE TRANSITION IN AMINO ACIDS

In 1991, in order to explain the origin of homochirality, Professor Abdus Salam suggested a possible phase transition in amino acids [6]. Starting from $Z^0$ interactions, he proposed that a quantum mechanical cooperative and condensation phenomena (possibly in terms of an e-n condensate where the e-n system has the same status as Cooper-pairing) give rise to a second-order phase transition. Based on BCS theory, gauge field theory and Higgs mechanism, he calculated the critical temperature for this phase transition. A crucial form for the transition temperature $T_c$ involves dynamical symmetry breaking. The idea is supposed to start from the Feynman Lagrangian methodology of the BCS theory for the superconducting electronic system. From Ginzburg-Landau equation, the value of temperature $T_c$ is deduced. By using Sakita's formulation, the following result is presented

$$T_C = \frac{\langle \varphi \rangle}{10^3} \exp[\frac{-2}{g_{eff}\sigma(1-4\sin^2\theta)}] \approx 2.5 \times 10^2 K$$

Here $\varphi$ field is expressed as a complex auxiliary Higgs scalar field, $g_{eff}$ is an attractive coupling constant between spin up and spin down electrons, $\theta$ is the Weinberg angle. Salam took *(1-4sin² θ) ≈ 1/13*, $\sigma(0)=m_z^2$ with the empirical value of the parameter *sin²θ ≈ 0.231*, and got *$g_{eff}\sigma(0) \approx 1$*.

In Salam's subsequent publications [7] (also some publication of Chela-Flores [8][9]), he predicted that this phase transition can eventually change D-alanine to L-alanine as $C_\alpha$-H bond would break and H atom become a superconductive atom. He had also proposed some experimental method to test his hypothesis: The first way to test for evidence is to lower the temperature while measuring the optical activity of a crystalline DL-alanine, D-alanine, L-alanine, when polarized light is shone upon a particular amino acid. If the polarization vector gets rotated, one may be sure that the appropriate phase transition has taken place. The second way of detecting the process may be by measuring differences of specific heats and looking for anomalies in the curve $C = \gamma T + \beta T^3$ +…like Mizutani have done for the non-amino acids like melanins and tumor melanosomes. Note that ideally Salam would be able to compute the values of $T_c$ when electroweak interaction is fully worked out. The analogy of the "superfluidity" exhibited by amino acids is to "superfluidity" in superconductors. In the case of superconductivity, the third way for a given amino acid is to apply an external magnetic field and look for the Meissner effect to determine $T_c$.

**EXPERIMENTAL EVIDENCE FOR A PHASE TRANSITION**

Following Salam's suggestion, our group at Peking University led by Professor Wang Wenqing has spent 8 years in searching for experimental evidence for Salam phase transition. A variety of experiments have been conducted to validate the existence of a phase transition[10][11][12].

*Experimental: Sample preparation/characterization*

The amino acids were obtained from Sigma Chemical Co. All single crystals were crystallized by slow evaporation of saturated aqueous solutions. D-/L-alanine and D-/L-valine single crystals were characterized by elemental analysis (C, H and N) and a good agreement was shown between the theoretical and experimental data [13]. By using X-ray diffraction crystallography at 293K, the cell dimensions of D-/ L-alanine crystals were determined as the same space group $P2_1P2_1P2_1$, orthorhombic, a = 6.0250 Å, b = 12.3310 Å, c = 5.7841 Å, V = 429.72 Å$^3$. The data for D-/L- valine crystals showed a space group $P2_1$, monoclinic, a= 9.6686 Å, b = 5.2556 Å, c = 11.9786 Å, V = 608.64 Å$^3$. It indicates that all crystals are pure single crystals containing no crystal water. The rotation angle $\zeta$ of the D- and L- alanine solution was measured on Polarimeter PE-241 MC at 293 K with the wavelength of 589.6 nm. By using the formula of $[\alpha] = \zeta / (L \times C)$, the corresponding $\alpha$ values of D- and L-alanine were shown to be the same.

In order to maintain the authenticity and accuracy of these experiments, we use dozens of pure single crystals for our series of experiments. We use different crystal in different experiment and all these results have been repeated.

*1.Specific heat measurement*

The temperature dependence of the specific heats for the D-, L- alanine and valine are shown in Fig .1-3 .An obvious $\lambda$ transition was observed at $270\pm1$ K in both alanine and valine enantiomers by differential scanning calorimetry with an adiabatic continuous heating method. With all the other conditions being equal, it is also shown that the specific heat $C_p$ value of D-valine is larger than that of L-valine and the same existed in the enantiomers of alanine. In all cases, the biologically dominant L-enantiomer is found to have lower energy, and the specific heat value reflects this fact.

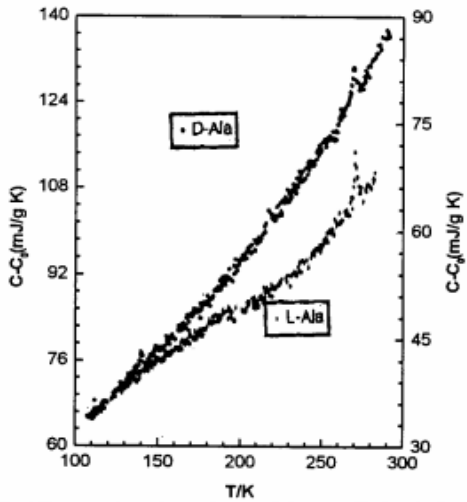
Fig.1 Temperature dependence of the specific heat of D-alanine, L-alanine

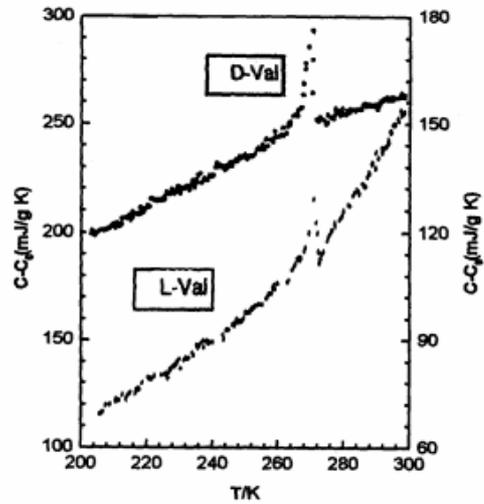
Fig.2 Temperature dependence of the specific heat of D-valine, L-valine

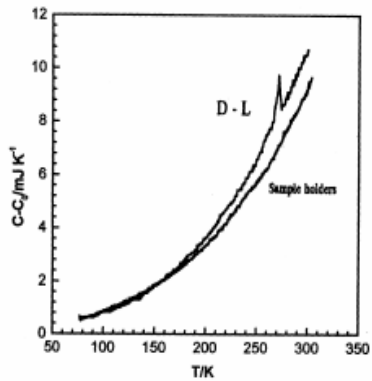
Fig.3 Temperature dependence of the differential specific heat between D-valine and L-valine

*2. DC-Magnetic susceptibilities measurements*

Magnetic moment ($m$) and magnetic susceptibility ($\chi_\rho$) of D-/ L-alanine crystals were measured by a SQUID magnetometer (Quantum Design, MPMS-5) from 200K to 300K at a field of 1.0$T$ (with differential sensitivity 1E-8 emu to 1 Tesla) in the National Laboratory for Superconductivity, Institute Physics Chinese Academy of Sciences. Crystals were weighed and determined to be 174.1mg (D-alanine), 99.5mg (L-alanine), then transferred to the straw. The signal from the plastic straw was canceled out while the temperature was measured.

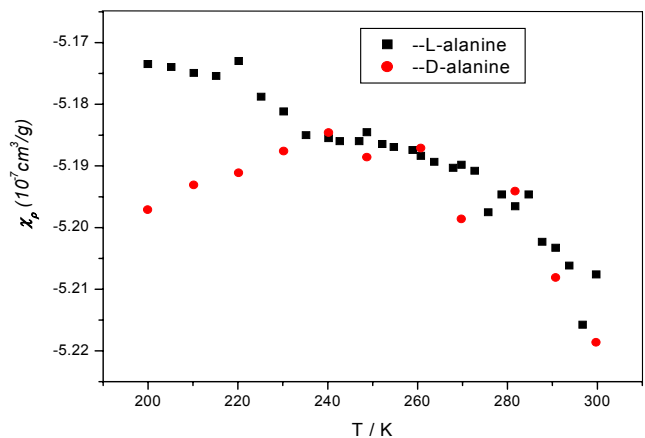

**Fig. 4 Temperature dependence of mass susceptibility at a field of 1.0 *T***

Magnetization measurements (shown in Fig.4) have shown alanine enantiomers to be dimagnetic, since they have even numbers of electrons which form closed magnetically neutral shells. It is clearly indicated that, both values and variations of magnetic susceptibilities of D-/ L- alanine enantiomers keep the same when the temperature is above 240K, and there is not distinct difference between them. It can also be found that, when temperature approaches 240K from higher ones, the change of $\chi_\rho$ values becomes slow and subsiding. On the other hand, when the temperature falls below 240K, D-alanine undergoes a magnetic phase transition as the $\chi_\rho$ values showed a maximum near 240K, while $\chi_\rho$ values of L -alanine go on increasing but the rate of variation experiences an abrupt rise, which should also be looked as a magnetic phase transition and its mechanism is clearly distinct with that of D-alanine crystal. The experimental results are repeatable for the same samples after several thermal circles from 200K to 300K, and both the magnitudes of $\chi_\rho$ values and the transition temperatures of D-alanine crystal are almost the same for the different thermal circles, which indicates that these changes are completely reversible for D- and L-alanine crystals.

Meaningfully, while the temperature varies in the range of 200~240K, $\chi_\rho$ values of D-/L-alanine are nearly opposite and symmetry with the $\chi_\rho$ value ($\chi_\rho = -5.185 \times 10^{-7}$ $cm^3/g$) of D-/L- alanine at 240K taken as the base value, which may show some clues for the weak neutral current playing a contrary even parity-violating role in magnetic behavior of molecules of D-alanine and L-alanine crystals in this temperature range.

*3.$^1$H CRAMPS solid state NMR measurements*

$^1$H NMR multipulse spectra were run on a Varian InfinityPlus–400 spectrometer with resonance frequency 400.12 MHz. A 4mm Chemagnetics double probe was used for the variable temperature CRAMPS experiment. A BR-24 multiple sequence was employed with a π/2 pulse width of 1.6 μs and 64 scans with a 2s recycle delay to acquire CRAMPS spectra. Spin rate was 2.5 kHz and the number of scans was 64. Chemical shifts were referenced to tetramethylsilane (TMS) for $^1$H measurements.

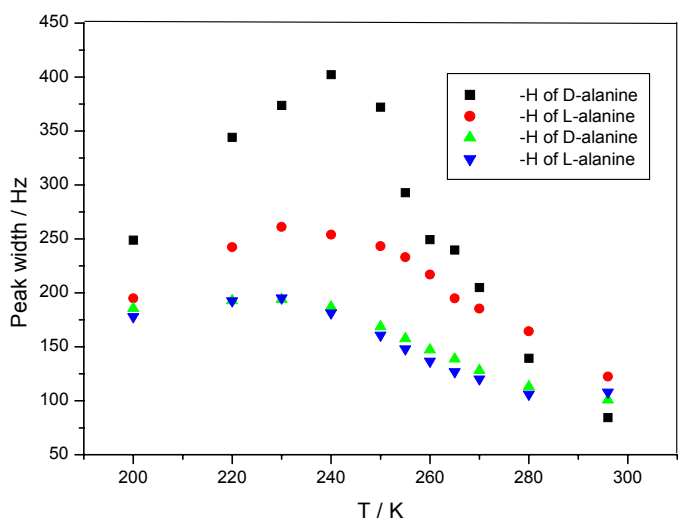

**Fig .5 Temperature dependence of peak widths of $α$-H, $β$-H of D-/L-alanine crystals**

For the sake of investigating the temperature-dependent nuclei dynamics of D- and L-alanine molecules, we measured the width of α-H and β-H peaks of D-/L-alanine. Fig.5 displays that four peak's width of *α*-H and *β*-H experience distinct maximum around 240K. So we conjecture that both D-/L-alanine may undergo a phase transition at this temperature range.

As for the L-alanine, the variation degrees of peak widths of $α$-H, $β$-H peaks make a good agreement (the maximum peak width is about two times as large as that in room temperature) in the whole process, which indicates that the temperature-dependent relaxation effects of $\alpha$-H, $\beta$-H nuclei of L-alanine molecule are nearly the same in the transition process. In the case of D-alanine, the variation of $\alpha$-H peak width is much fiercer than that of its enantiomer in the transition temperature range (220~250K), in addition, its transition temperature seems nearer to 240K instead of 230K. Considering the relative stability of magnetic field in the whole experimental process, these results show that, in this specific transition, the spin-spin relaxation and spin-lattice relaxation mechanism of $\alpha$-H nucleus of D-alanine molecule may be different from that of L-alanine, and its relaxation effects may also stronger than its enantiomer.

*4. Ultrasonic attenuation measurements*

The measurements of ultrasonic attenuation values of D-/L-alanine were performed on a computer controlled ultrasonic system (Matec model 7700). A ceramic transducer for generating a longitudinal ultrasonic wave was used. The frequency of ultrasonic wave is 5.56MHz. The attenuation α value was calculated from

$$\alpha = \frac{1}{d} \ln \left[ \frac{A(x_1)}{A(x_2)} \right],$$

where $d$ is the thickness of the sample, $A(x_1)$ and $A(x_2)$ are the amplitudes of the first and second echoes respectively. The samples used for ultrasonic measurements were D- and L-alanine crystals with thickness equal to 3.00 and 2.70 mm. The transducer was fixed properly to ensure the contact with the sample and there was no air between them. The temperature of the sample was controlled by a thermocouple. In this way we can gain the ultrasonic attenuation values at various temperatures. For the sake of detecting and describing the difference of both phase transitions mechanisms more precisely, relative attenuation values curve for either enantiomer was drew, respectively: the maximum attenuation value of either enantiomer was chosen as $\alpha_{max}$, the other temperature-dependent $\alpha$ values were normalized by dividing them with $\alpha_{max}$. The experiments were performed for the cooling process from 290K to 200K.

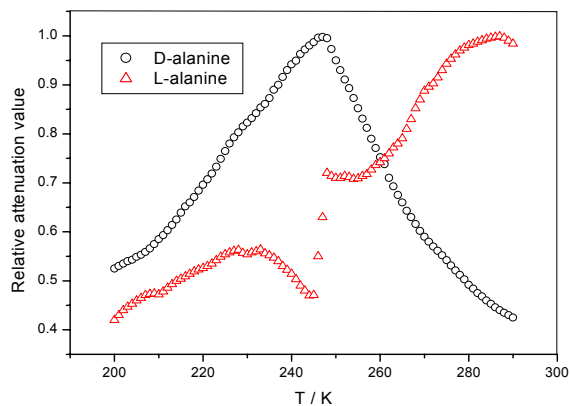

**Fig.6 The relative attenuation values of D-/L-alanine crystals vs its temperature**

The results of ultrasonic measurements are shown in Figure.6. As for the D-alanine crystal, the attenuation value changes continuously and undergoes an obvious peak around 247K, and the curve is symmetry on the whole. In the case of L-alanine crystal, a precipitous step-form drop of attenuation value was found around 247K. These ultrasonic attenuation phase transition temperatures agree with the $^1$H Solid state NMR and DC-magnetic susceptibilities measurements. It is quite clear that, for these two crystal samples, the shapes of relative attenuation curves of the enantiomers and transition mechanisms are quite different from each other. So we suppose the ultrasonic investigations demonstrate the contribution of the parity-violating WNC to the elastic relaxation effects acting in both ultrasonic transitions of D-alanine and L-alanine crystals.

*5. Temperature-dependent optical rotation measurement*

This is the first method suggested by Salam to testify the phase transition. The temperature-dependent optical rotation angle result of DL racemic alanine was performed on a solid state optical measurement system and the result was shown in Fig.7. The $\varphi$ value is equal to $-0.5°$ (approach to zero) from 290K to 252K. It proves that the crystal is truly racemic. However, when the temperature continuously decreases from 260K pass through 252K to 230K, the $\varphi$ value is rapidly decreasing to $-4.5°$ showing an obvious characteristic phase transition. The experimental results are repeatable over several thermal circles from 230K through to 290K, which indicates that these behaviors are completely reversible for DL- alanine crystal. For different DL- alanine crystal samples, we observed the same temperature-dependent $\varphi$ behavior characterized by approximately the same transition temperature.

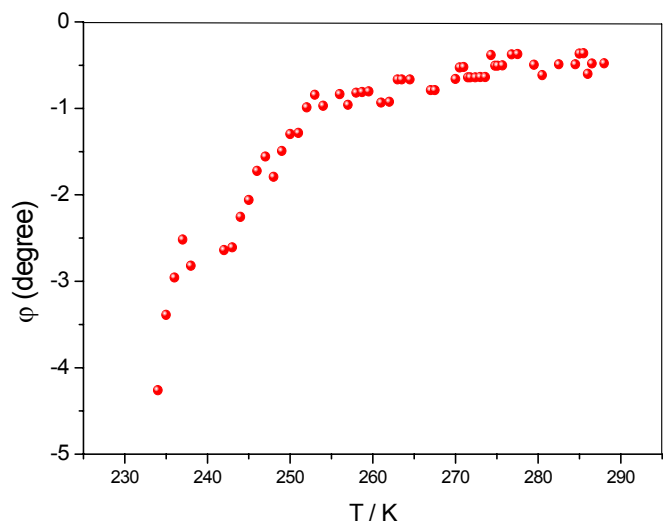

**Fig. 7 Temperature dependent optical rotation angle in DL racemic crystal**

*6. Temperature-dependent Raman studies*

Vibration spectra of DL-alanine single crystals, which were measured in above optical rotation angle experiments, were performed on a Ranishaw 1000 fitted with argon-ion laser model. The temperature was monitored from 290K to 230K by a digital temperature controller unit having a stability of 0.1K. As in the optical rotation angle measurements, the laser beam also propagated along the c axis perpendicular to the {110} face. The experimental conditions were as follows: exciting line, 514.5nm, and power, 50mW; the width of slit of the triple monochromator, 50 $\mu$m; scanning range, 100-3600cm$^{-1}$; signal averaging, 3 scans.

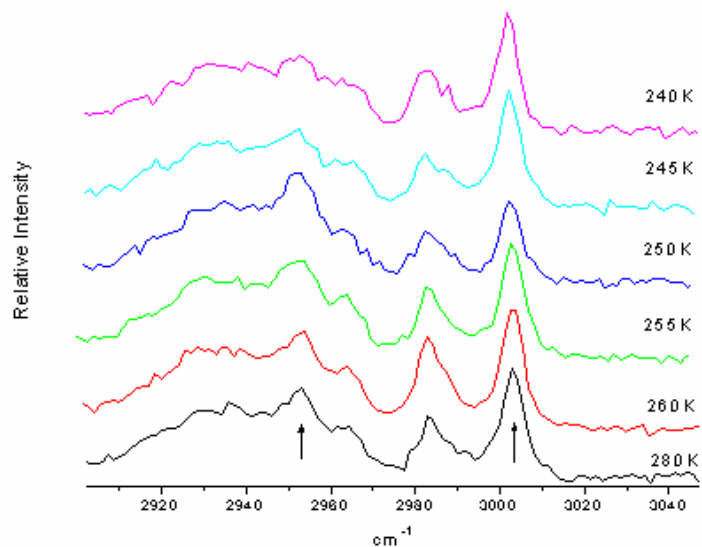

Fig. 8 Temperature-dependent Raman spectra of DL-alanine in the region from 2920 cm-1 to 3040 cm-1

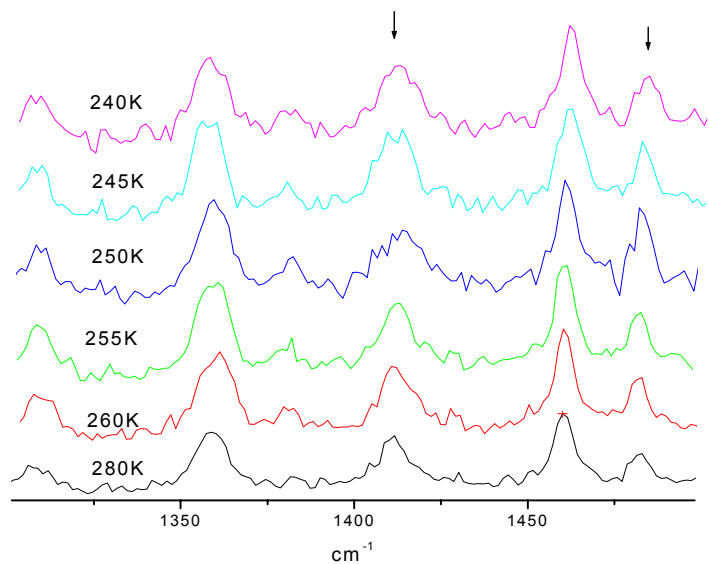

**Fig. 9 Temperature-dependent Raman spectra of DL-alanine crystal in the region from 1300 cm$^{-1}$ to 1500 cm$^{-1}$**

The most striking and considerably important result of this part of study is the apparent different Raman spectra behavior of DL racemic alanine crystal when temperature approaches the transition temperature $T_c$. A careful observation of this phase transition by Raman spectra showed that (according to the assignment of peaks in reference ): (1) the relative intensity of 3002 cm$^{-1}$ ($C_\alpha$–H stretching) is weaker than that of 2953 cm$^{-1}$ only at 250K (shown in Fig. 8), (2) the relative peak intensity of 1411 cm$^{-1}$ peak (COO$^-$ symmetric bending) is weaker than that of 1483 cm$^{-1}$ ($C_\alpha$–H bending) only at 250K in the whole temperature region (shown in Fig. 9). These phenomena imply the abrupt change of vibration particulars of $C_\alpha$–H bond and COO$^-$ group at the transition point around 250K, which suggests the contribution of $C_\alpha$–H bond and COO$^-$ group made to the phase transition process of DL- alanine crystal.

*Conclusion*

From above series of experimental results, the existence of a phase transition around 250K in alanine crystals has been fully proved. Although the detailed transition mechanism is unclear, we may at least draw a cautious conclusion that there exists a phase transition in alanine crystals and in the transition process, D-, L-, and DL- alanine crystals exhibit different transition behavior.

**EXPERIMENTAL EVIDENCE THAT DENY THE CONFIGURATION CHANGE**

In 1993, Figureau et al. conducted a series of scrupulously performed experiments to test the validity of Salam's predictions [14][15]. They observed no change in optical rotation after exposing both racemic DL-cystine and L-cystine to temperature ranging from 77K to 0.6K for three and four days, thus reported failing to validate PVED-induced phase transitions predicted by Salam.

In addition to their experiments, a more direct way to testify Salam's phase transition is to conduct temperature-dependent X-ray diffraction or neutron diffraction on alanine enantiomers. If there is a configuration change of D- to L-, it will be easy to catch this phenomenon for an abrupt change in atom coordinates will be observed.

**TABLE 1. Temperature dependent X-ray diffraction data for D-/L- alanine**

| Temperature | 300K | |
|---|---|---|
| Samples | D-Alanine | L-Alanine |

| | | | | | | | |
|---|---|---|---|---|---|---|---|
| Empirical formula | | $C_3H_7NO_2$ | | | | | |
| Formula weight | | 89.10 | | | | | |
| Crystal system | | Orthorhombic | | | | | |
| Space group | | | | $P2_12_12_1$ | | | |
| Unit cell Dimensions (nm) | | a=0.6088(1) | | | | 0.60344(5) | |
| | | b=1.2267(3) | | | | 1.23668(8) | |
| | | c=0.5800(2) | | | | 0.57879(3) | |
| Volume (nm³) | | V=0.4331(2) | | | | 0.43193(5) | |
| Z | | | 4 | | | | |
| Atomic coordinates | | X | Y | Z | X | Y | Z |
| O(1) | | 0.7281(2) | 0.08438(7) | 0.3716(1) | 0.7282(1) | 0.08437(6) | 0.3716(1) |
| O(2) | | 0.4505(2) | 0.18519(8) | 0.2390(1) | 0.4503(1) | 0.18512(7) | 0.2388(1) |
| C(1) | | 0.4764(2) | 0.1611(1) | 0.6444(2) | 0.4766(2) | 0.16118(8) | 0.6444(2) |
| C(2) | | 0.5604(2) | 0.14163(8) | 0.3975(2) | 0.5606(2) | 0.14163(7) | 0.3978(2) |
| H(4) | | 0.441(2) | 0.238(1) | 0.656(3) | 0.440(2) | 0.237(1) | 0.655(3) |
| Dihedral angle θ | | | 49.53 ° | | | 49.40 ° | |

| | |
|---|---|
| Temperature | 270K |

| | | | | | | | |
|---|---|---|---|---|---|---|---|
| Samples | | D-Alanine | | | L-Alanine | | |
| Empirical formula | | $C_3H_7NO_2$ | | | | | |
| Formula weight | | 89.10 | | | | | |
| Crystal system | | Orthorhombic | | | | | |
| Space group | | | | $P2_12_12_1$ | | | |
| Unit cell Dimensions (nm) | | a =0. 60073(5) | | | | 0.60095(5) | |
| | | b=1.23030(7) | | | | 1.23388(7) | |
| | | c = 0.57732(4) | | | | 0.57904(3) | |
| Volume (nm³) | | V=0.42669(5) | | | | 0.42936(5) | |
| Z | | | 4 | | | | |
| Atomic coordinates | | X | Y | Z | X | Y | Z |
| O(1) | | 0.7280(1) | 0.08419(6) | 0.3723(1) | 0.7279(1) | 0.08414(6) | 0.3725(1) |
| O(2) | | 0.4488(1) | 0.18497(6) | 0.2388(1) | 0.4489(1) | 0.18494(6) | 0.2388(1) |
| C(1) | | 0.4744(2) | 0.16118(8) | 0.6442(2) | 0.4746(2) | 0.16115(8) | 0.6445(2) |
| C(2) | | 0.5591(2) | 0.14146(7) | 0.3978(1) | 0.5593(2) | 0.14146(7) | 0.3980(2) |
| H(4) | | 0.441(2) | 0.237(1) | 0.656(2) | 0.440(2) | 0.238(1) | 0.656(2) |
| Dihedral angle θ | | | 45.52 ° | | | 45.20 ° | |

| | |
|---|---|
| Temperature | 250K |

| | | |
|---|---|---|
| Samples | D-Alanine | L-Alanine |
| Empirical formula | $C_3H_7NO_2$ | |
| Formula weight | 89.10 | |

| Crystal system | Orthorhombic | | | | | |
|---|---|---|---|---|---|---|
| Space group | | $P2_12_12_1$ | | | | |
| Unit cell Dimensions (nm) | a=0.60041(3) | | | | 0.60002(4) | |
| | b=1.23179(5) | | | | 1.23293(6) | |
| | c=0.57892(3) | | | | 0.57822(2) | |
| Volume (nm$^3$) | V=0.42815(3) | | | | 0.42775(4) | |
| Z | | 4 | | | | |
| Atomic coordinates | X | Y | Z | X | Y | Z |
| O(1) | 0.7282(1) | 0.08412(6) | 0.3727(1) | 0.7279(1) | 0.08404(6) | 0.3730(1) |
| O(2) | 0.4481(1) | 0.18482(6) | 0.2388(1) | 0.4480(1) | 0.18484(6) | 0.2387(1) |
| C(1) | 0.4734(2) | 0.16125(8) | 0.6443(2) | 0.4736(2) | 0.16115(8) | 0.6447(2) |
| C(2) | 0.5582(2) | 0.14134(7) | 0.3980(1) | 0.5587(2) | 0.14139(7) | 0.3984(1) |
| H(4) | 0.437(2) | 0.237(1) | 0.653(2) | 0.440(2) | 0.237(1) | 0.658(2) |
| Dihedral angle θ | | **43.97°** | | | **45.72°** | |

The data for D-/L- alanine single crystals are listed in Table 1, which are collected at different temperatures on a Siemens R3m/V diffractometer with $M_o$--$K_x$ ($\lambda$=0.71073Å) radiation. The structure is solved by direct methods and subsequent Fourier Differential Techniques, and refined by full-Matrix least squares using the SHELXTL PLUS program. All non-hydrogen atoms are refined with anisotropic thermal parameters.

Dihedral angles was calculated from the atomic coordinates of O(1) O(2) C(1) C(2) H(1) of D- and L-alanine under the temperature dependence of X-ray diffraction data. The result of temperature-dependent X-ray studies shows no evidence for structural changes in this temperature range.

## NEW VIEWS ON SALAM PHASE TRANSITION

It seems a contradiction to admit the existence of Salam phase transition but deny the configuration change of D- to L-. All those experimental evidence suggest a revision of Salam hypothesis. We may divide Salam hypothesis into two parts: 1. Theoretical deduction of a specific phase transition based on quantum mechanics and BCS theory. 2. His prediction of a configuration change of D- to L- during the phase transition process. All the experiments that refute Salam hypothesis are actually against the second part of his hypothesis not the first part. A better definition of Salam phase transition should be: a phase transition in which D- and L- enantiomers exhibit different transition behavior.

Although Salam's hypothesis has failed to explain homochirality only via this phase transition, the significance of the different behavior between D- and L- amino acids in the phase transition will never be ignored. The parity violation effect in weak interaction has made enantiomers different. Berger and Quack's study in a detailed analysis of dynamic chirality proved that the dihedral angle between the $O_2C$ and $C_\alpha$-H planes plays an important role in determining the intrinsic energies of the alanine molecules and this difference has been used in the calculation of Parity Violating Energy Difference (PVED) [17]. According to Quack's theoretical method by means of highest level ab initio studies (MC-LR), the PVED value is $1.2 \times 10^{-19}$ Hartree ($3.3 \times 10^{-18}$ eV/molecule), namely L-alanine is more stable than D-alanine. D-alanine and L-alanine have shown obvious difference at 250K. The most striking result that confirm this study in our X-ray diffraction is the apparent difference in the dihedral angle between D- and L- alanine at 250K (D:43.97°  L:45.72°) while in other temperature they are almost the same. Quack has also proved that the difference between enantiomers induced by weak neutral currents is too minor to be detected by experimental methods under normal conditions [18]. But as we have already illustrated, D-, L-, and DL- alanine crystals display distinct behavior in all experimental methods in the specific phase transition process. A possible explanation could be: the minor difference between D- and L- enatiomers (PVED: about $10^{-18}$ - $10^{-17}$ eV) has been enlarged to a detective level during this phase transition due to the quantum mechanical cooperative and condensation phenomena. This amplified energy difference will become the foundation of later amplification mechanism.

Establishing chemical systems that are thermodynamically far from equilibrium, Kondepudi and Nelson succeeded in simulating the evolution of dominant chirality by nonlinear analysis of chemical reactions. If we take the specific phase transition into consideration, the enlarged energy difference will greatly shorten the time period needed in their model to achieve homochirality and it will also reduce the total number of molecules that this system must have in order to surmount fluctuation. What is worth emphasizing, according to Salam, the occurrence of Salam phase transition can enhance the PVED a lot, which will in turn induce much higher probability of transition. This kind of nonlinear feedback relation is similar with that in Kondepudi-Nelson scenario : when the PVED value is larger by one to two orders of magnitude than anticipated, the time required to realize homochirality in a 100km×100km×4m lake reduces from $10^4$ year to one year[19][20]. In fact, nonlinear effects in asymmetric synthesis and stereoselective reactions have been widely discovered and studied [21][22]. A recently published experiment showed that the chiral amplification of oligopeptides in two-dimensional crystalline self-assemblies on water also introduces nonlinear effects [23].

Significantly, the biological system when homochirality was achieved meets well the formation requirements of a dissipative structure — the critical transition point is far from equilibrium; the control parameter PVED, whose value determines the probability of Salam phase transition, has a transition threshold; the occurrence of Salam phase transition will in turn enlarge PVED nonlinearly. Hence, if the temperature is kept in the transition temperature range all along, the control parameter PVED can exceed its critical value (threshold) for enough time, and it is possible to form a dissipative structure and finally develop into an ordered and homochiral system. This scenario is being tested in our laboratory. A related and interesting phenomenon was reported by Ellis-Evans *et al*[24]: a great lake—Lake Vostok lies below a flowing ice sheet, and the melting point temperature is –3.15 ºC (270K). Microbiological studies of the Vostok ice core have revealed a great diversity of microbes including yeasts and actinomycetes (with antibiotic synthesizing potential) which remain viable in ice for up to 3000 years, and viable mycelial fungi up to 38600 years old.

At last, we emphasize again the significance of Salam phase transition in the evolution of homochirality: instead of the ultimate solution to the problem, it may actually plays as the first step of amplification mechanism. It connects the microcosmic difference (PVED) between biomolecular enantiomers with nonlinear process in a macrocosmic biological system. It also solves the long-term debating suspicion that PVED is too minor to be enlarged directly by nonlinear process. Combining the existence of PVED, Salam phase transition and nonlinear amplification mechanism, we may propose a sound way to understand the chemical evolution of homochirality as depicted in Fig.12.

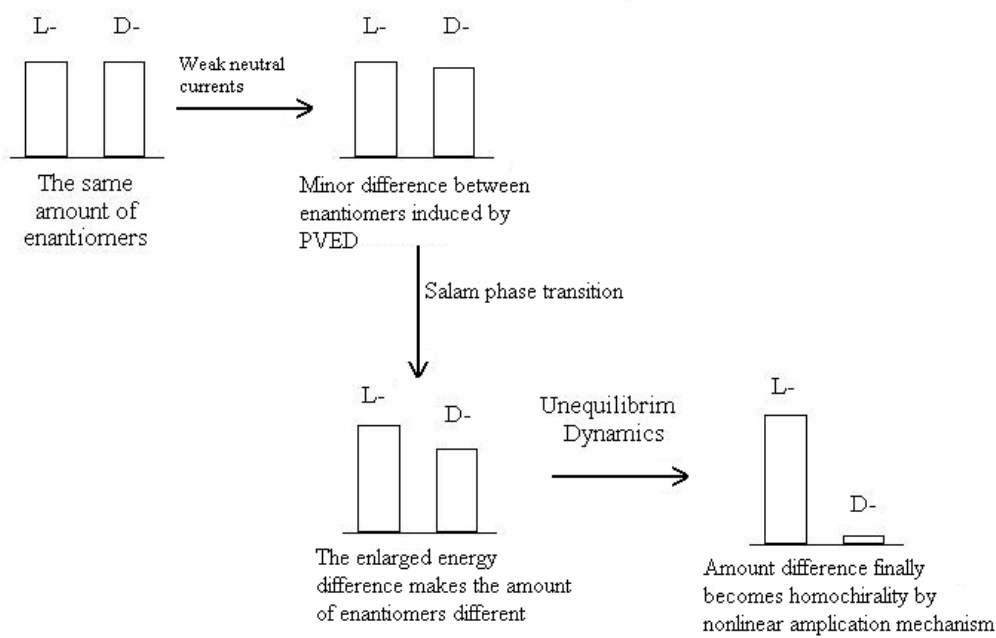

**Fig. 12 A possible evolution process of homochirality**

**Acknowledgments**

This research was supported by the grant of 863 program (863-103-13-06-01) and by the grant of National Natural Science Foundation of China (29672003).